\documentclass[10pt,fleqn]{article}
\usepackage{amsfonts,amssymb}
\usepackage{amsmath} 
\allowdisplaybreaks[1]

\let\kappa=\varkappa

\usepackage[text={17cm,24cm},centering]{geometry}

\newcommand{\dd}{\;\mathrm{d}}
\newcommand{\deriv}[2]{\mbox{$\displaystyle \frac{{\rm d}#1}{{\rm d}#2}$}}
\newcommand{\pderiv}[2]{\,\mbox{$\displaystyle\dfrac{{\partial}#1}{{\partial}#2}$}\,}

\newcommand{\gn}{\gamma_{\mathrm{N}}}
\newcommand{\np}{\phi_{\mathrm{N}}}
\newcommand{\bm}[1]{\boldsymbol{#1}}
\newcommand{\abc}[1]{\mbox{#1)}\quad}
\newcommand{\pot}[1]{\cdot 10^{#1}}
\newcommand{\kk}[1]{\,\mathrm{#1}}

\newlength{\BAR}
\newcommand{\newbar}[1]{\settowidth{\BAR}{$#1$}%
\stackrel{\rule{0.8\BAR}{0.03em}}{\,\!#1}{}}
\newcommand{\nb}[1]{\newbar{#1}}
\newcommand{\vS}{\varSigma}

\newlength{\temp}

\newcounter{saveeqn}
\newcommand{\alpheqn}{ \setcounter{saveeqn}{\value{equation}}%
\addtocounter{saveeqn}{1}
\setcounter{equation}{0}%
\renewcommand{\theequation}{\mbox{\arabic{saveeqn}\alph{equation}}}}
\newcommand{\reseteqn}{\setcounter{equation}{\value{saveeqn}}%
\renewcommand{\theequation}{\arabic{equation}}}

\newcommand{\ci}[1]{\stackrel{\circ}{#1}{\!\!}}
\newcommand{\vP}{\varPhi}
\newcommand{\vO}{\varOmega}
\newcommand{\PM}{\genfrac{}{}{0pt}{1}{+}{(-)}\,}

\newcommand{\Deriv}[2]{\deriv{#1_{#2}}{t}}

\newenvironment{Literatur}{\begin{list}{}{\parsep 2pt plus1pt minus1pt
\itemsep \parsep \leftmargin4em\itemindent-4em}}{\end{list}}

\begin{document}
\title{%
Investigation of the 2-body system with a rotating central body (e.
g. earth-moon system) within the Projective Unified Field theory:
the transfer of rotational angular momentum and energy from the
central body to the orbital 2-body system, the tidal and the
non-tidal influences (mechanical, general-relativistic
Lense-Thirring effect and cosmological PUFT-contributions)}

\author{        E.Schmutzer\thanks{eschmu@aol.com}, Jena, Germany  \\
         Friedrich Schiller University Jena}
\date{}

\maketitle

\begin{abstract}
In this treatise the well-known 2-body problem with a rotating
central body is systematically reinvestigated on the basis of the
Projective Unified Field Theory (PUFT) under the following aspects
(including the special case of the Newton mechanics): First,
equation of motion with abstract additional terms being appropriate
for the interpretation of the various effects under discussion:
tidal friction effect as well as non-tidal effects (e.g. rebound
effect as temporal variation of the moment of inertia of the
rotating body, general-relativistic Lense-Thirring effect, new
scalaric effects of cosmological origin, being an outcome of the
scalarity phenomenon of matter (PUFT). Second,  numerical evaluation
of the theory.\\[1ex]
\textsc{Key words}: two-body problem with rotating central body --
tidal and non-tidal effects -- scalaric-cosmological influence of
the expanding cosmos on the 2-body system.
\end{abstract}

\section{Basis of the new approach}\label{sec:1}

Caused by the enormous progress in astrophysical precision measuring
technique during the last decade (Williams et al. 2003, Anderson et
al. 2002, Schneider 2003; further references in these papers),
mainly based on the satellite tracking technique, especially the
satellite laser ranging, the theoretical interest in this field of
research grew considerably. Obviously the main part of the transfer
of rotational angular momentum from the rotating earth to the
orbital angular momentum of the whole earth-moon system has its
origin in the tidal braking of the mentioned rotation by the
gravitational interaction between the orbiting moon and the flowing
viscous waters of the oceans (Lambeck 1980, Brosche and Schuh 1998,
Sabadini and Vermeersen 2004; further references in these
publications).

Nowadays particular empirical investigations refer to the additional
non-tidal effects, since because of the high precision measuring
technique, the latter can empirically be separated from the tidal
effects. In this context some further fields of research are
investigated now: Apart from the post-glacial rebound effect already
mentioned in the above quoted monograph of Lambeck and in greater
detail investigated in a recent publication (Wu, Schuh and Bibo
2003, further references in this paper), in the general-relativistic
Einstein mechanics the change of the gravitational field outside a
rotating body, particularly the ``gravito-magnetic'' contribution of
the Lense-Thirring effect should be mentioned (L\"ammerzahl and
Neugebauer 2001, where further basic references can be found).
Beyond that the current theoretical and empirical investigations aim
to various other scientific subjects. Our present field-theoretical
research on the basis of PUFT points to new scalaric-cosmological
results being also interesting for application to the earth-moon
system (Schmutzer 2004).

\subsection{Equation of motion of the 2-body system}

We start with our equation of motion for the case of
non-relativistic velocities and weak gravitational and scalaric
fields (reference mentioned above) and apply it to two different
bodies. Then as usual we pass over to the combined single equation
of motion for the 2-body problem. Having in mind the later
application to the 2-body system being under cosmological influence,
we arrive at the following equation of motion for such a mechanical
system (dot means derivative with respect to the time $t$):
\begin{align}\label{1}
   \abc{b} \ddot{\bm{r}}+\frac{\gn M}{r^3}\,\bm{r}+\varSigma \dot{\bm{r}}+
    \bm{g}_{az|\mathrm{rad}}=0\quad\text{with}\qquad\abc{b}
    \varSigma=\frac{1}{\newbar{\sigma}}\,\deriv{\newbar{\sigma}}{t}\,,
\end{align}
where $\newbar{\sigma}$  is the scalaric world function and  $\vS$
is the ``logarithmic  scalaric world function'', both (of
non-Newtonian origin) describing the cosmological influence on the
2-body system. Further $\bm{g}_{az|\mathrm{rad}}$ is  a free
abstract azimuthal-radial acceleration term introduced for grasping
the transfer mechanism of rotational angular momentum from the
rotating central body to the orbital angular momentum of the 2-body
system.

We remember that (apart from the two additional terms mentioned) the
equation (\ref{1}) refers to the frame of reference of the center of
mass of both bodies. As mentioned above, our treatment of the 2-body
system is based on an orbiting body 1 (because of  thinking of the
moon later index $M$) and an orbiting body 2 (because of  thinking
of the earth later index $E$), both bodies orbiting about the common
center of mass. Of course, the theory being developed here can be
applied to appropriate 2-body systems occurring in astrophysics in
general.

Let us further mention that $\gn$ is the Newtonian gravitational
constant as a true constant of nature. Since nowadays there exist
serious hints for a time dependence of the empirical gravitational
constant $G( t)$, we have strictly to distinguish between these two
quantities.

In order to make the reading of this paper easier, we mainly use the
notation of our treatment of the 2-body system in our textbook
(Schmutzer 1989, 2005).\\[1ex]
List of notations:
\alpheqn
    \begin{align}\label{2a}
    &m_{M}           &&  \text{(mass of the moon),}\\
\label{2b}
    & m_{E} &&  \text{(mass of the earth),}\\
\label{2c}
    &M=m_E+m_M &&  \text{(total mass of the system),}\\
\label{2d}
    &m_{red}=\frac{m_E m_M}{M}&&  \text{(reduced mass);}\\[1ex]
\refstepcounter{saveeqn}\label{3a}\setcounter{equation}0
    &\bm{r}_E=-\frac{m_M}{M}\,\bm{r}&&\text{(radius vector from
the center of mass to the earth),}\\
\label{3b}
    &\bm{r}_M=\frac{m_E}{M}\,\bm{r}&&\text{(radius vector from the center of mass to the
    moon),}\\
\label{3c}
    &\bm{r}=\bm{r}_M-\bm{r}_E\,,\\
\label{3d}
    &r=|\bm{r}|\,.
\end{align}
\reseteqn
Further following relation is valid:
\begin{align}\label{4}
    m_E\bm{r}_E+m_M\bm{r}_M=0\,.
\end{align}
\subsection{Conservation of the angular momentum of the 2-body
system}

The total angular momentum of the 2-body system reads:
\begin{align}\label{5}
    \bm{L}_{tot}=\bm{L}_{orb}+\bm{L}_{rot}\,,
\end{align}
where
\begin{align}\label{6}
    \bm{L}_{orb}=m_{red}\,\bm{r}\times\dot{\bm{r}}
\end{align}
is the orbital angular momentum of the 2-body system and
\begin{align}\label{7}
    \abc{a}\bm{L}_{rot}=\bm{k}L_{rot}\quad \text{with}\quad
    \abc{b}L_{rot}=\theta_{E}\omega_{E}
\end{align}
is the rotational angular momentum of the rotating central body
($\theta_{E}$ moment of inertia, $\omega_{E}$ angular velocity,
$\bm{k}$ constant unit vector orthogonal to the $x$-$y$-plane).

For physical reasons it seems to be acceptable to demand the
conservation law of the total angular momentum of the 2-body system:
\begin{align}\label{8}
    \deriv{\bm{L}_{tot}}{t}=0\,.
\end{align}
But one should realize that separate conservation laws do not hold:
\begin{align}\label{9}
    \abc{a}
    \deriv{\bm{L}_{rot}}{t}=\bm{k}\deriv{{L}_{rot}}{t}\quad\text{with}\quad
    \abc{b}\deriv{{L}_{rot}}{t}=\theta_{E}\dot{\omega}_{E}+\dot{\theta}_{E}\omega_{E}\,,
\end{align}
\begin{align}\label{10}
    \deriv{\bm{L}_{orb}}{t}=\bm{M}_{az|rad}\,,
\end{align}
where $\bm{M}_{az|rad}$ is the torque acting on the 2-body system.

In order to treat the orbital angular momentum of the 2-body system,
we first cast the equation of motion (\ref{1}) into the appropriate
form
\begin{align}\label{11}
    m_{red}\,\ddot{\bm{r}}+\bm{e}_{r}\frac{\gn
    Mm_{red}}{r^2}+m_{red}\vS\dot{\bm{r}}+m_{red}\bm{g}_{az|rad}=0\,.
\end{align}
From the equation of motion (\ref{11}) we find in the usual way the
following expression for the torque:
\begin{align}\label{12}
    \bm{M}_{az|rad}=-m_{red}\,\bm{r}\times\bm{g}_{az|rad}\,.
\end{align}
Since the torque and the torque causing force $\bm{F}_{az|red}$ are
related to each other by the formula
\begin{align}\label{13}
    \bm{M}_{az|rad}=\bm{r}\times\bm{F}_{az|rad}\,,
\end{align}
we obtain by comparing this result to the alternative result
(\ref{12}) the relationship
\begin{align}\label{14}
    \bm{F}_{az|rad}=-m_{red}\bm{g}_{az|rad}\,.
\end{align}
With the help of (\ref{12}) the balance equation (\ref{10}) takes
the shape
\begin{align}\label{15}
    \deriv{\bm{L}_{orb}}{t}=-m_{red}\,\bm{r}\times\bm{g}_{az|rad}\,.
\end{align}

\subsection{Balance of the mechanical energy of the 2-body system}

In contrast to the demand of conservation of the total angular
momentum of the 2-body system, because of the irreversible processes
(e.g. friction of viscous waters of the oceans of the earth) here we
are confronted with a non-conservation phenomenon of the mechanical
energy. Therefore we have to balance the whole energy content of the
2-body system.

Let us now first define the total mechanical energy of the 2-body
system considered:
\begin{align}\label{16}
    E_{tot}=E_{orb|grav}+T_{rot}\,,
\end{align}
where the orbital-gravitational energy is defined by
\begin{align}\label{17}
    E_{orb|grav}=T_{orb}+U_{grav}
\end{align}
with
\begin{align}\label{18}
   \abc{a}T_{orb}=\frac{1}{2}\,m_{red}\,\dot{\bm{r}}{}^{2}&&\text{(orbital kinetic
   energy),}&&\abc{b}
        U_{grav}=-\frac{\gn M m_{red}}{r}&&\text{(gravitational energy).}
\end{align}
The rotational kinetic energy reads
\begin{align}\label{19}
    T_{rot}=\frac{1}{2}\,\theta_{E}\omega_{E}^2\,.
\end{align}
By differentiation from (\ref{16}) results
\begin{align}\label{20}
    \deriv{E_{tot}}{t}=\deriv{E_{orb|grav}}{t}+\deriv{T_{rot}}{t}\,.
\end{align}
From these calculations we obtain the following formula for the
orbital-gravitational energy of the 2-body system, being used in
this explicit shape later:
\begin{align}\label{21}
    E_{orb|grav}=\frac{1}{2}\,m_{red}\dot{\bm{r}}{}^{2}-\frac{\gn M
    m_{red}}{r}\,.
\end{align}
Now it is necessary to make some assumptions on the time dependence
of several mass quantities.

In our PUFT (1995) we introduced the concept that the usual mass of
a body (being for cosmological reasons in principle time-dependent)
is the product of two factors:
\begin{align}\label{22}
    m(t)=\mathcal{M}\nb{\sigma}(t)\,,
\end{align}
where $\mathcal{M}$ is the time-independent scalmass of this
considered body as the primary essential mass (Urmasse) being
modified by the scalaric world function $\nb{\sigma}$  of
cosmological origin (see Schmutzer 2004). If this concept is
accepted, then for $m(t)$ results
\begin{align}\label{23}
    \dot{m}=m\vS\,.
\end{align}
Application of this concept to the mass quantities introduced above
leads to following formulas:
\begin{align}\label{24}
    &\abc{a}\dot{m}_{E}=m_{E}\vS\,,&&\abc{b}\dot{m}_{M}=m_{M}\vS\,,&&\abc{c}
    \dot{M}=M\vS\,,&&\abc{d}\dot{m}_{red}=m_{red}\vS\,.
\end{align}
With respect to the moment of inertia we are confronted with a
particular situation: Admitting an additional, empirically motivated
explicit time dependence of this quantity, we arrive at the relation
\begin{align}\label{25}
    \abc{a}\dot{\theta}_{E}=\theta_E\vS+\ci{\theta}_E\quad\text{with}\quad\abc{b}
\ci{\theta}_E=\pderiv{\theta_E}{t}\,.
\end{align}
For the following it is suitable to split the additional
acceleration $\bm{g}_{az|rad}$ occurring in the equation (\ref{1})
into an azimuthal and a radial term, using the polar coordinates
$\bigl\{R,\vP\bigr\}$ in the $x$-$y$-plane:
\begin{align}\label{26}
    \bm{g}_{az|rad}=\bm{e}_{R}g_{rad}+\bm{e}_{\vP}g_{az}\,.
\end{align}
Now using the results (\ref{23}) and (\ref{25}), after rather
lengthy calculations we find the following expression for the time
derivative of the orbital-gravitational energy:
\begin{align}\notag
    \deriv{E_{orb|grav}}{t}&=-\Bigl(\frac{1}{2}\,m_{red}\dot{\bm{r}}{}^{2}+
    \frac{2\gn M m_{red}}{r}\Bigr)\vS-
    m_{red}\bigl(R\dot{\vP}g_{az}+\dot{R}g_{grav}\bigr)\\ \label{27}
    &=-\bigl(T_{orb}-2U_{grav}\bigr)\vS-m_{red}\bigl(R\dot{\vP}g_{az}+\dot{R}g_{rad}\bigr)\,.
\end{align}
If we apply the abbreviations
\begin{align}\label{28}
    \abc{a}\deriv{E_{az|rad}}{t}=m_{red}\bigl(R\dot{\vP}g_{az}+\dot{R}g_{rad}\bigr)\,,\qquad
    \abc{b}\deriv{Q_{s}}{t}=-\bigl(T_{orb}-2U_{grav}\bigr)\vS\,,
\end{align}
the expression (\ref{26}) reads
\begin{align}\label{29}
 \deriv{E_{orb|grav}}{t}= \deriv{Q_s}{t}- \deriv{E_{az|rad}}{t}\,.
\end{align}
With respect to the rotating central body, we defined its rotational
kinetic energy already above in the usual way (\ref{19}):
\begin{align}\label{30}
    T_{rot}=\frac{1}{2}\,\theta_{E}\omega_{E}{}^{2}\,.
\end{align}
Hence with the help of (\ref{25}) for the time derivative results
\begin{align}\label{31}
    \deriv{T_{rot}}{t}=\deriv{Q_{fric}}{t}+\frac{1}{2}\,\dot{\theta}_{E}{\omega}_{E}{}^2
    =\deriv{Q_{fric}}{t}+T_{rot}\vS +\frac{1}{2}\,
    \ci{\theta}_{E}\omega_{E}^2\,,
\end{align}
where the abbreviation
\begin{align}\label{32}
    \deriv{Q_{fric}}{t}=\theta_E \omega_E\dot{\omega}_{E}<0
\end{align}
for the irreversible heat production of the rotating central body
(being braked by the internal friction of the viscous fluid
component of this body) was used. Let us mention that braking means
fulfilling the condition $\dot{\omega}_E<0$.

Inserting of (\ref{27}) and (\ref{31}) into (\ref{20}) leads by
means of (\ref{28}) to the final formula
\begin{align}\label{33}\notag
    \deriv{E_{tot}}{t}&=\deriv{Q_{s}}{t}+\deriv{T_{rot}}{t}-\deriv{E_{az|rad}}{t}
    \\ &=
    \deriv{Q_{fric}}{t}+\frac{1}{2}\,
   { \ci{\theta}}_{E}\omega_E^2-\bigl(T_{orb}-T_{rot}-2U_{grav}\bigr)\vS-
    m_{red}\bigl(R\dot{\vP}g_{az}+\dot{R}g_{rad\bigr)\,.}
\end{align}

\section{Transition of the equation of motion to cylindrical
coordinates}\label{sec:2}

\subsection{Decomposition of the equation of motion}

For practical application of the general theory presented in section
\ref{sec:1} it is favorable to use the cylindrical coordinates
$\bigl\{R,\vP,z\bigr\}$.

Then the equation of motion (\ref{1}) splits into
\begin{subequations}\label{34}
\begin{align}\label{34a}
& \ddot{R}-R\dot{\vP}{}^2+\frac{\gn M
R}{\bigl(R^2+z^2\bigr)^{3/2}}+\dot{R}\vS+g_{rad}=0 &&\text{(radial
equation),}\\ \label{34b}
&R\ddot{\vP}+2\dot{R}\dot{\vP}+\dot{\vP}R\vS+g_{az}=0 &&
\text{(azimuthal equation),}\\
\label{34c}&\ddot{z}+\vS\dot{z}+\frac{\gn
Mz}{\bigl(R^2+z^2\bigr)^{3/2}}=0
    &&  \text{(axial equation).}
\end{align}
\end{subequations}
For special use in astrophysics it is appropriate to derive by
differentiation of (\ref{34b}) the intermediate result
\begin{align}\label{35}
    \ddot{R}=\frac{1}{2}\,\biggl(\frac{\ddot{\vP}}{\dot{\vP}}\biggr)^2R-
    \frac{1}{2\dot{\vP}}\,\bigl(\dddot{\vP}R+\ddot{\vP}\dot{R}\bigl)-\frac{1}{2}\bigl(\dot{R}\vS
    +\dot{\vS}R\bigr)-\frac{1}{2\dot{\vP}}\,\dot{g}_{az}+\frac{1}{2}\,\frac{\ddot{\vP}}{\dot{\vP}^2}\,
    g_{az}\,.
\end{align}
Inserting into (\ref{34a}) leads to
\begin{align}\label{36}
    R^{2}\dot{\vP}^2-3\dot{R}^2-\frac{\gn
    M}{R}=-\frac{R^2}{2\dot{\vP}}\,\dddot{\vP}-\frac{R^2}{2}\,(\dot{\vS}-\vS^2)
    +3R\dot{R}\vS+\frac{g_{az}}{2\dot{\vP}}\,\bigl(\vS
    R+3\dot{R}\bigr)-\frac{R}{2\dot{\vP}}\,\dot{g}_{az}+Rg_{rad}\,.
\end{align}

\subsection{General annotation on the adiabaticity approximation}

Since the exact integration of the very complicated system of
non-linear differential equations (\ref{34}) is hopeless, we use in
the following our adiabaticity approximation introduced in our
monograph (Schmutzer 2004).

In astrophysics/cosmology we have to consider the time dependence of
physical quantities on two different levels of consideration:
\begin{itemize}
\item[--]  quick motion of bodies, particles etc. (relatively quick
compared to the cosmological          temporal changes),
\item[--]
cosmological motion (slow time dependence of the cosmological
quantities).
\end{itemize}

Practically the adiabaticity approximation means that the
integrations are performed in such a way that the slow cosmological
time dependence is neglected, i.e. that during the integration the
time plays a non affected parameter role.

\subsection{Integration of the axial equation}

This integration is performed under the conditions
\begin{align}\label{37}
    \abc{a}
    \frac{z}{R}\ll 1\,,\quad \abc{b}R\approx R_0
\end{align}
(near to the $x$-$y$-plane and near to circularity). Then from
(\ref{34}c) the well-known oscillation equation
\begin{align}\label{38}
    \ddot{z}+\vS\dot{z}+\frac{\gn M}{R_0^3}\, z=0
\end{align}
results. Using the same abbreviations as in our textbook (Schmutzer
1989, 2005):
\begin{align}\label{39}
    \abc{a}\varrho=\frac{1}{2}\, \vS\,,&&
    \abc{b}\omega_0=\PM\sqrt{\frac{\gn M}{R^3_0}}\,,&&\abc{c}
        \omega_z=\sqrt{\omega_0^2-\varrho^2}=\sqrt{\frac{\gn
    M}{R_0^3}-\frac{1}{4}\,\vS^2}\,,
\end{align}
and referring to the small friction case ($\varrho<\omega_0$), then
the solution reads ($A$, $B$ constants of integration)
\begin{align}\label{40}
    z=e^{-\frac{1}{2}\,\vS t}\bigl[A\sin(\omega_zt)+\cos(\omega_z
    t)\bigr]\,.
\end{align}
The exponential factor describes the cosmologically caused damping
of the axial motion towards the $x$-$y$-plane with the damping
parameter $\varrho$. Further $\omega_z$ means the angular frequency
of the orbital motion. The corresponding revolution time is
\begin{align}\label{41}
    \tau_z=\frac{2\pi}{\omega_z}\,,
\end{align}
whereas for the logarithmic decrement of damping follows
\begin{align}\label{42}
\delta_z=\ln \frac{z_n}{z_{n+1}}=\frac{1}{2}\,\tau_z\vS\,.
\end{align}

\subsection{Radial and azimuthal equations of motion in the plane of
motion}

In this case both the equations (\ref{34a}) and (\ref{34b}) for $z =
0$ read:
    \begin{align}\label{43}
   &\abc{a}\ddot{R}-R\vO^{2}+\frac{\gn M}{R^2}+\vS\dot{R}+g_{rad}=0&&
    \text{or}&&\abc{b}
    g_{rad}=-\ddot{R}+R\vO^2-\frac{\gn
    M}{R^2}-\dot{R}\vS\,;\\
    \label{44}
    &\abc{a}R\dot{\vO}+(2\dot{R}+R\vS)\vO+g_{az}=0&&\text{or}&&\abc{b}
        g_{az}=-(R\dot{\vO}+2\vO\dot{R}+R\vO\vS)\,,
\end{align}
where the abbreviation $\vO=\dot{\vP}$ was used. These equations
will be exploited later.

\section{Treatment of the conservation of the angular momentum in the
plane of motion}\label{sec:3}

\subsection{Orbital angular momentum}

In specialization to cylindrical coordinates the orbital angular
momentum takes the form
\begin{align}\label{45}
    \abc{a}\bm{L}_{orb}=\bm{k}L_{orb}\,,\quad\text{where}\quad\abc{b}
    L_{orb}=m_{red}F=m_{red}R^2\vO\quad\text{with}\quad\abc{c}F=R^2\vO\,.
\end{align}
One should remember that in Newton mechanics in the case of
vanishing external influence on the orbiting system the quantity $F$
(equivalent to the angular momentum) is a constant of motion, but
here, as we know from (\ref{10}), because of the rotating central
body, $L_{orb}$ is not a constant of motion.

Differentiation of (\ref{45}a) leads to the formula
\begin{align}\label{46}
    \deriv{\bm{L}_{orb}}{t}=\bm{k}\deriv{L_{orb}}{t}\,.
\end{align}
Comparison to (\ref{10}) gives
\begin{align}\label{47}
    \bm{M}_{az|rad}=\bm{k}M_{az|rad}=\bm{k}\deriv{L_{orb}}{t}\,,\quad\text{i.e.}
\end{align}
for the corresponding component the relation
\begin{align}\label{48}
    M_{az|rad}=\deriv{L_{orb}}{t}
\end{align}
holds.

Our next step consists  in differentiating of (\ref{45}b) and
eliminating in the obtained result by means of (\ref{44}a). We find
\begin{align}\label{49}
    \deriv{L_{orb}}{t}=-m_{red}\,Rg_{az}\,.
\end{align}
Substituting this expression into (\ref{48}) leads to
\begin{align}\label{50}
    M_{az|rad}\rightarrow M_{az}=-m_{red}R g_{az}\,.
\end{align}
This result coincides with formula (\ref{12}), being obvious if we
use the relation (\ref{26}) for elimination.

From (50) we see that $g_{rad}$ does not enter into the torque, i.e.
the torque is determined by $g_{az}$ only.

Remembering the general relationship between the torque and the
torque causing force we arrive at the formulas:
\begin{align}\label{51}
    \abc{a}\bm{F}_{az|rad}\rightarrow
    \bm{F}_{az}=\bm{e}_{\vP}F_{az}\quad\text{with}\quad\abc{b}
    F_{az}=-m_{red}g_{az}\,.
\end{align}
Let us remind the reader that we already above calculated the
quantity $g_{az}$ (\ref{44}b).

\subsection{Rotational angular momentum and the conservation law of
the total angular momentum}

The rotational angular momentum and its time derivative are
presented in the formulas (\ref{7}) and (\ref{9}). Next we write
(\ref{5}) by means of (\ref{45}a) and (\ref{7}a) in the form
\begin{align}\label{52}
    \bm{L}_{tot}=\bm{k}L_{tot}\quad\text{with}\quad \abc{b}
    L_{tot}=L_{orb}+L_{rot}\,.
\end{align}
Now we insert (\ref{49}) and (\ref{9}b) into the conservation law of
the total angular momentum (\ref{8}) in the form
\begin{align}\label{53}
    \deriv{L_{tot}}{t}=\deriv{L_{orb}}{t}+\deriv{L_{rot}}{t}=0
\end{align}
and arrive at the relation
\begin{align}\label{54}
    \dot{\omega}_{E}=-\frac{m_{red}R^2\vO}{\theta_E}\,\Bigl(\frac{\dot{\vO}}{\vO}+2\frac{\dot{R}}{R}+\vS\Bigr)
    -\omega_E\vS
    -\frac{1}{\theta_{E}}\,\ci{\theta}_{E}\omega_E
\end{align}
or rearranged at
\begin{align}\label{55}
    \vS=-\frac{1}{m_{red}R^2\vO+\theta_E\omega_E}\,\biggl[\theta_E\dot{\omega}_E+m_{red}R^2\vO\Bigl(\frac{\dot{\vO}}{\vO}
    +2\frac{\dot{R}}{R}\Bigr)+\ci{\theta}_{E}\omega_E\biggr]\,,
\end{align}

\section{Treatment in the plane of motion: energy balancing, temporal
change of the orbital-gravitational energy, total angular momentum}

\subsection{Balancing of the total energy}

Our calculations in section \ref{sec:1} arrived at the balance
equation for the total energy of the system considered (\ref{33}).
Further the orbital kinetic energy (\ref{18}a) and the gravitational
energy of the 2-body system (\ref{18}b) read in cylindrical
coordinates as follows:
\begin{align}\label{56}
    \abc{a}T_{orb}=\frac{1}{2}\,m_{red}\bigl(\dot{R}^{2}+R^2\vO^2\bigr)\,,
    \quad
    \abc{b}
    U_{grav}=-\frac{\gn M m_{red}}{R}\,.
\end{align}
Hence instead of (\ref{21}) for the plane of motion follows
\begin{align}\label{57}
    E_{orb|grav}=T_{orb}+U_{grav}=\frac{1}{2}\,m_{red}\bigl(\dot{R}^{2}+R^2\vO^2\bigr)-
    \frac{\gn M m_{red}}{R}\,.
\end{align}
Let us now remember the formulas (\ref{43}b) for $g_{rad}$,
containing $\ddot{R}$ (2nd order derivative of $R$), and (\ref{44}b)
for $g_{az}$ , containing only first order derivates of the involved
physical quantities being interesting for us. Up till now we don't
have any empirical information on the quantity $\ddot{R}$.
Therefore, of course we have to think about how to get some
knowledge on this interesting radial second order effect. Here
obviously on the way via the quantity $g_{rad}$, i.e. by getting
further information on this quantity.

For this reason we cast the balance equation (\ref{33}) into the
form
\begin{align}\label{58}
    \Deriv{E}{tot}=\Deriv{Q}{fric}+E_{tot}\vS\,.
\end{align}
The fulfillment of this postulate is reached by the choice
\begin{align}\label{59}
    g_{rad}=\Bigl(2+\frac{R\dot{\vO}}{\vO
    \dot{R}}\Bigr)R\vO^2-\frac{1}{\dot{R}}\Bigl(\dot{R}^2+\frac{\gn
    M}{R}\Bigr)\vS+\frac{1}{2m_{red}}\,\ci{\theta}_E\omega_E^2\,.
\end{align}
Hence the radial equation (\ref{43}a) takes the form
\begin{align}\label{60}
    \ddot{R}+R\vO^2+R^2\vO\frac{\dot{\vO}}{\dot{R}}+\frac{\gn
    M}{R^{2}}\,
    \Bigl(1-\frac{R}{\dot{R}}\,\vS\Bigr)+\frac{1}{m_{red}\dot{R}}\,\ci{\theta}_{E}\omega^2_{E}
    =0\,.
\end{align}
Future precision experiments on radial second order effects could
show whether this equation is applicable.

Further we derive with the help of (\ref{54}) from (\ref{44}b)
\begin{align}\label{61}
    g_{az}=\frac{\theta_E}{R
    m_{red}}\Bigl(\dot{\omega}_{E}+\omega_E\vS+\frac{1}{\theta_E}\,\ci{\theta}_{E}\omega_{E}\Bigr)
\end{align}
and from (\ref{28}a) the result
\begin{align}\label{62}
    \Deriv{E}{az|rad}=-m_{red}\Bigl(R^2\vO^2+\dot{R}^2+\frac{\gn
    M}{R}\Bigr)\vS+\frac{1}{2}\,\ci{\theta}_E\omega^2_E\,.
\end{align}
Then (\ref{29}) goes over to the simpler formula
\begin{align}\label{63}
    \abc{a}
    \Deriv{E}{orb|grav}=E_{orb|grav}\vS-\frac{1}{2}\,\ci{\theta}_{E}\omega^2_E
    \quad\text{or}\quad\abc{b}
    \deriv{}{t}\biggl(\frac{E_{orb|grav}}{m_{red}}\biggr)=-\frac{1}{2m_{red}}\,
    \ci{\theta}_E\omega^2_E\,.
\end{align}

\subsection{Temporal change of the orbital-gravitational energy}

Then integration of (\ref{63}b) leads to the following expression
for the time-dependent orbital-gravitational energy:
\begin{align}\label{64}
    \abc{a}
    E_{orb|grav}=-m_{red}\Lambda+\frac{1}{2}\,\hat{C}m_{red}
    \quad\text{with}\quad
    \abc{b}\Lambda=\frac{1}{2}\int
    \frac{\ci{\theta}_E\omega^2_E}{m_{red}}\dd t\quad\text{($\hat{C}$
    constant of integration).}
\end{align}
We explicitly learn from these relations that primarily the origin
of the increase of the orbital-gravitational energy of the 2-body
system considered is caused by the time-dependent decrease of the
moment of inertia of the rotating body (rebound effect).

Next by means of  (\ref{64}a) we derive from (\ref{57}) the equation
\begin{align}\label{65}
    \abc{a}\dot{R}^2+R^2\vO^2-\frac{2\gn M}{R}=\hat{C}-2\Lambda\,,
    \quad\text{i.e.}\quad
    \abc{b}
    \dot{R}=\PM\sqrt{\hat{C}+\frac{2\gn M}{R}-R^2\vO^2-2\Lambda}\,.
\end{align}

\subsection{Explicit conservation law of the total angular momentum}

In order to treat this task we give the equation (\ref{54}) by means
of (\ref{25}a) the shape
\begin{align}\label{66}
    \abc{a}
    m_{red}\deriv{}{t}\bigl(R^2\vO\nb{\sigma}\bigr)+
    \theta_E\deriv{}{t}\bigl(\omega_E\nb{\sigma}\bigr)+\ci{\theta}_E\omega_E\nb{\sigma}=
    0\quad\text{or}\quad
    \abc{b}
    \deriv{}{t}\bigl(m_{red}R^2\vO+\theta_E\omega_E\bigr)=0\,.
\end{align}
Integration leads to the conservation law of the total angular
momentum
\begin{align}\label{67}
    J_{0}=L_{tot}=L_{orb}+L_{rot}=m_{red}R^2\vO+\theta_E\omega_E
\end{align}
($J_{0}$  constant of integration).

\section{Transition to the Keplerian conic motion by using the
adiabaticity approximation}\label{sec:5}

Eliminating in (\ref{65}a) the angular velocity $\vO$ derived from
(\ref{67}) in the form
\begin{align}\label{68}
    \vO=\frac{J_{0}-\theta_E\omega_E}{m_{red}R^2}\,,
\end{align}
we arrive at the equation
\begin{align}\label{69}
    \dot{R}^2+\frac{(J_0-\theta_E\omega_E)^2}{m_{red}^2R^2}-\frac{2\gn
    M}{R}=\hat{C}-2\Lambda
\end{align}
to be treated further now. Since in the parameters $M$, $m_{red}$,
$\theta_E$, $\omega_E$, $\Lambda$ being involved in this non-linear
differential equation a weak adiabatic (partly
adiabatic-cosmological)  time dependence occurs, we approximately
take them as adiabatic constants, when we apply the transformation
\begin{align}\label{70}
    R(t)=\frac{1}{u(t)}
\end{align}
in order to reach our goal of the adiabatic Keplerian conic equation
of motion ($\varepsilon$ excentricity)
\begin{align}\label{71}
    \abc{a}
    \deriv{^2u}{t^2}+u-\frac{\nb{A}}{\varepsilon}=0\quad\text{with}
    \quad\abc{b}
\frac{\nb{A}}{\varepsilon}=\frac{\gn M
m_{red}^2}{(J_0-\theta_E\omega_E)^2}\,.
\end{align}
The solution of (\ref{71}a) reads in an appropriate form
\begin{align}\label{72}
    R=\frac{\varepsilon}{\nb{A}(1-\varepsilon\cos\vP)}\,.
\end{align}
In the following we are only interested in the elliptic motion which
according to the notion in our monograph (Schmutzer 1989, 2005) is
determined by ($\bar{a}$, $\bar{b}$ semi-axes)
\begin{align}\label{73}
    \varepsilon=\sqrt{\bar{a}{}^2-\bar{b}{}^2}<1
    \qquad(\bar{b}<\bar{a})\,.
\end{align}
Calculating the energy of the 2-body system by inserting of
(\ref{68}), (\ref{72}) etc. into (\ref{57}) leads to the result
\begin{align}\label{74}
    E_{orb|grav}=-\frac{\gn M m_{red}}{2\bar{a}}
\end{align}
which formally reminds us of the situation in the Keplerian 1-body
system, but one should not forget the adiabatic time dependence in
the physical quantities occurring in (74).

Next we calculate from the adiabatically valid formula for the area
of the ellipse
\begin{align}\label{75}
    A_{ell}=\frac{\pi
    \varepsilon^2}{\nb{A}^2(1-\varepsilon^2)^{3/2}}
\end{align}
and from the also adiabatically valid Keplerian area law
\begin{align}\label{76}
    \Deriv{A}{ell}=\frac{A_{ell}}{\tau_{ell}}=\frac{J_0-\theta_E\omega_E}{2m_{red}}
\end{align}
the revolution time $\tau_{ell}$ of the elliptic motion:
\begin{align}\label{77}
    \tau_{ell}=\frac{2\pi}{\sqrt{\gn M}}\,a^{2/3}\,.
\end{align}
Using the angular revolution frequency
\begin{align}\label{78}
    \vO_{ell}=\frac{2\pi}{\tau_{ell}}\,,
\end{align}
we find the differential relationship
\begin{align}\label{79}
    \frac{\dd \vO_{ell}}{\vO_{ell}}=-\frac{3\dd a}{2 a}
\end{align}
which because of (\ref{74}) is valid for the change  of $\vO_{ell}$,
understandable through the change of the orbital-gravitational
energy $E_{orb|grav}$, caused by external energetic influences. In
this context we remember that without such external influences
(apart from the weak cosmological effects mentioned) and without a
temporal change of $\theta_E$ the orbital-gravitational energy
according to (\ref{63}) would be conserved.

Let us here remind that the relation (\ref{79}) should not be
confused with the relation
\begin{align}\label{80}
    \frac{\dd \vO}{\vO}=-2\frac{\dd R}{R}
\end{align}
which follows from the conservation law of the angular moment
($R^2\vO=const$) of the non-disturbed Keplerian 1-body system. See
equation (45b) for the case $F = const$.

\section{Influence of the Lense-Thirring effect on the 2-body
system}\label{sec:6}

It is well-known that the Newtonian gravitational field equation is
at most usable in the quasi-static case of time dependence of the
source. Already in the case of a stationarily rotating source the
Einsteinian gravitational theory leads to a modification of the
Newtonian external field which was calculated in first approximation
by J.~Lense and H.~Thirring (1918). Nearly half a century later R.
P.~Kerr succeeded in finding an exact exterior solution of a
rotating body.

It is obvious that the Lense-Thirring modification of the
gravitational field soon became interesting also for
astrophysicists. Parallel to the field modification also the
modification of the corresponding equation of motion of a body was
derived by these authors, leading to the Lense-Thirring additional
terms in the extended Newtonian equation of motion
\begin{align}\label{81}
    \ddot{\bm{r}}+\bm{e}_r\frac{\gn M}{r^2}-\frac{2\gn}{c^2r^3}\,
    \dot{\bm{r}}\times\bigl[\bm{L}_E-3(\bm{L}_E\bm{e}_r)\bm{e}_r\bigr]
    =0
\end{align}
($\bm{L}_E=\bm{k}\,\theta_E\omega_E$ angular momentum of the
rotating central body, $c$ velocity of light). A new derivation of
this equation was recently published (L\"ammerzahl and Neugebauer
2001, further references in this paper).

In analogy to the Coriolis acceleration
\begin{align}\label{82}
    \bm{a}_{C}=2\dot{\bm{r}}'\times\bm{w}
\end{align}
one is inclined to define
\begin{align}\label{83}
    \bm{w}_{\mathrm{LTh}}=\frac{\gn}{c^2r^3}\,\bigl[\bm{L}_E-3(\bm{L}_E\bm{e}_r)\bm{e}_r\bigr]
\end{align}
as the Lense-Thirring frame dragging  angular velocity.

With the aim to compare the equation (\ref{81}) with our ansatz
equation (\ref{1}) we on the basis of the Newton mechanics ($\vS=0$)
specialize the equations (\ref{81}) and (\ref{1}) to the plane of
motion ($\bm{e}_R$ and $\bm{e}_{\vP}$ usual unit vectors in this
plane) and find by using of (\ref{26}):
\begin{subequations}\label{84}
\begin{align}\label{84a}
    &\ddot{\bm{R}}+\bm{e}_R\frac{\gn}{R^2}\,\Bigl(M-\frac{2\theta_E\omega_E\vO}{c^2}\Bigr)
    +\bm{e}_{\vP}\frac{2\gn \theta_E\omega_E\dot{R}}{c^2R^3}=0\,,\\
    \label{84b}
    &\ddot{\bm{R}}+\bm{e}_R\frac{\gn
    M}{R^2}+\bm{e}_Rg_{rad}+\bm{e}_{\vP}g_{az}=0\,.
\end{align}
\end{subequations}
Comparing these two results, we arrive at the identifications
\begin{align}\label{85}
    \abc{a}g_{rad|\mathrm{LTh}}=-\frac{2\gn
    \theta_E\omega_E\vO}{c^2 R^2}\,,\quad\abc{b}
g_{az|\mathrm{LTh}}=\frac{2\gn
    \theta_E\omega_E\dot{R}}{c^2 R^3}\,.
\end{align}
In this special case ($\vS=0$) from (\ref{59}) and (\ref{44}b)
follows:
\begin{align}\label{86}
    \abc{a}g_{rad}=\Bigl(2+\frac{R\dot{\vO}}{\vO\dot{R}}\Bigr)R\vO^2+
    \frac{1}{2m_{red}}\,\ci{\theta}_E\omega^2_E
    \,, \quad\abc{b}
    g_{az}=-(R\dot{\vO}+2\vO\dot{R})\,.
\end{align}
Of course, the pairs of both the equations (\ref{85}) and (\ref{86})
have a fully different physical basis.

Numerical evaluation shows that in the case of the earth-moon system
the order of magnitude of the general-relativistic Lense-Thirring
effects (\ref{85}) are roughly 14 orders smaller than the above
calculated effects. Nevertheless it may be interesting to look how
for appropriate star models (yet to look for) the situation could
change.

\section{Survey of the exterior spherical harmonics expansion of the
Newtonian    gravitational potential including various applications
up to the second order}\label{sec:7}

\subsection{General mass distribution}

On the basis of the Newtonian gravitational field equation
\begin{align}\label{87}
    \triangle\np =4\pi\gn \mu
\end{align}
($\np$ Newton potential, $\mu$ mass density of the source) the
expansion considered reads (Lambeck 1980, Schneider 1999, Schmutzer
1989,2005):
\begin{align}\label{88}
    \np(r,\vartheta,\varphi)=-\frac{\gn M}{r}
    \biggl[1+\sum\limits_{n=1}^{\infty}\sum\limits_{m=0}^{n}\biggl(\frac{r_e}{r}\biggr)^n
    \,P^{m}_{n}(\cos\vartheta)
    \biggl\{C_{nm}\cos{m\varphi}+S_{nm}\sin(m\varphi)\biggr\}\biggr]\,,
\end{align}
where following notation is used: \settowidth{\temp}{Stokes
coefficients (tesseral gravitational moments, geopotential
harmonic}
\begin{align}\label{89}
    \begin{aligned}
      &M&&\parbox[t]{\temp}{mass of the source body,}\\
&r_{e} &&\parbox[t]{\temp}{radial parameter,}\\
&\varphi &&\parbox[t]{\temp}{geographic length,}\\
&\vartheta    &&\parbox[t]{\temp}{polar angle
($\beta=\dfrac{\pi}{2}+\vartheta$ geographic latitude),}\\
&C_{nm}\,,\ S_{nm} &&\parbox[t]{\temp}{Stokes coefficients (tesseral
gravitational moments, geopotential harmonic
coefficients),}\\
    &P^m_n  &&\parbox[t]{\temp}{associated Legendre
    polynomials.}
    \end{aligned}
\end{align}
Choosing the center of mass as origin of the spherical coordinate
system, one finds the zero values:
\begin{align}\label{90}
    C_{10}=0\,,\quad C_{11}=0\,,\quad S_{11}=0\,.
\end{align}
Then the equation (\ref{88}) reads
\begin{multline}\label{91}
    \np(r,\vartheta,\varphi)=-\frac{\gn M}{r}
    \biggl[1-\sum\limits_{n=2}^{\infty}\biggl(\frac{r_e}{r}\biggr)^n
    J_{n}P_{n}(\cos\vartheta)\\+
    \sum\limits_{n=2}^{\infty}\sum\limits_{m=1}^{n}\biggl(\frac{r_e}{r}\biggr)^n
    \,P^{m}_{n}(\cos\vartheta)
    \biggl\{C_{nm}\cos({m\varphi})+S_{nm}\sin(m\varphi)\biggr\}\biggr]\,,
\end{multline}
where the abbreviation
\begin{align}\label{92}
    J_n=-C_{n0}\qquad\text{(zonal gravitational moments)}
\end{align}
is used.

\subsection{General mass distribution in second order approximation}

In this approximation, presupposed for our further investigations,
the dynamic form factor (quadrupole moment parameter)
\begin{align}\label{93}
    J_2=-C_{20}
\end{align}
is of  our specific interest.

For some further calculations it is convenient to go over to
Cartesian coordinates (Greek indices run from 1 to 3). Then the
Newton potential takes the form
\begin{align}\label{94}
    \np(\bm{r})=\np^{(M)}(\bm{r})+\np^{(Q)}(\bm{r})
\end{align}
with
\begin{subequations}\label{95}
\begin{align}\label{95a}
&\np^{(M)}(\bm{r})=-\frac{\gn M}{r} &&\text{(monopole potential),
}\\
&\np^{(D)}(\bm{r})=0 && \text{(vanishing dipole potential),}\\
&\np^{(Q)}(\bm{r})=-\frac{\gn}{2r^5}\bigl(3M_{\alpha\beta}x_\alpha
x_\beta-M_{\alpha\alpha}r^2\bigr)&& \text{(quadrupole potential),}
\end{align}
\end{subequations}
where
\begin{subequations}\label{96}
\begin{align}\label{96a}
&M=\int\limits_V\mu(\bar{\bm{r}})\,\dd \nb{V}      &&\text{(mass of
the
source),} \\
&M_{\alpha\beta}=M_{\beta\alpha}=\int\limits_V\mu(\bar{\bm{r}})\bar{x}_\alpha
\bar{x}_\beta\,\dd \nb{V} &&\text{(mass tensor of the source).}
\end{align}
\end{subequations}

By means of this quantity the mass quadrupole moment tensor
\begin{align}\label{97}
    \abc{a}
    m_{\alpha\beta}=m_{\beta\alpha}=3M_{\alpha\beta}-M_{\gamma\gamma}\delta_{\alpha\beta}
    \quad\text{with}\quad
    \abc{b}
    m_{\alpha\alpha}=0\,,\quad
    \abc{c}M_{\alpha\beta}=\frac{1}{3}\,(m_{\alpha\beta}+M_{\gamma\gamma}\delta_{\alpha\beta})
\end{align}
is defined. Then the potential (\ref{94}) with the help of
(\ref{95}a), (\ref{95}c) and (\ref{97}) reads
\begin{align}\label{98}
    \np(\bm{r})=-\frac{\gn
    M}{r}-\frac{\gn}{2r^5}\,m_{\alpha\beta}x_{\alpha}x_{\beta}\,.
\end{align}

\subsection{Inertia moment tensor and mass quadrupole moment tensor}

The inertia moment tensor
\begin{align}\label{99}
    \theta_{\alpha\beta}=\theta_{\beta\alpha}=\int\limits_V\mu(\bar{\bm{r}})\bigl(
    \bar{x}_\gamma\bar{x}_\gamma\delta_{\alpha\beta}-\bar{x}_\alpha\bar{x}_\beta\bigr)
    \,\dd\nb{V}
\end{align}
which is related to the mass tensor (\ref{96}b) and to the mass
quadrupole moment tensor (\ref{97}a) via the relations
\begin{align}\label{100}
    \abc{a}\theta_{\alpha\beta}=M_{\gamma\gamma}\delta_{\alpha\beta}-M_{\alpha\beta}
    \quad\text{with}\quad\abc{b}\theta_{\gamma\gamma}=2M_{\gamma\gamma}
\end{align}
and
\begin{align}\label{101}
    \theta_{\alpha\beta}=\frac{1}{3}\bigl(\theta_{\gamma\gamma}\delta_{\alpha\beta}-m_{\alpha\beta}\bigr)
    \,.
\end{align}
In main-axes representation of the inertia moment tensor the
following relationships with the Stokes coefficients hold (Schneider
1999):
\begin{align}\label{102}
    &\abc{a}
    C_{20}=-J_2=-\frac{\theta_{33}-\frac{1}{2}\bigl(\theta_{11}+\theta_{22}\bigr)}{Mr_{e}^2}\,,
    &&\abc{b}
    C_{21}=-\frac{\theta_{13}}{Mr_{e}^2}=0\,,
    &&\abc{c}C_{22}=-\frac{\theta_{11}-\theta_{22}}{4Mr^2_e}\,;\\
    \label{103}
    &\abc{a}S_{21}=-\frac{\theta_{23}}{Mr_e^2}=0\,,
    &&\abc{b}S_{22}=-\frac{\theta_{12}}{2Mr^2_e}=0\,.
\end{align}
Introducing for simplicity the usual notation
\begin{align}\label{104}
    \abc{a}
    \theta_{11}=\theta_x=A\,,\quad\abc{b}\theta_{22}=\theta_y=B\,,
    \quad
    \abc{c}\theta_{33}=\theta_z=C
\end{align}
($A$, $B$  equatorial moments of inertia, $C$ polar moment of
inertia), we arrive at the following shape of both the non-vanishing
coefficients:
\begin{align}\label{105}
 \abc{a}   J_{2}=\frac{C-\frac{1}{2}\,(A+B)}{Mr_e^2}\,,
 \quad\abc{b}C_{22}=\frac{B-A}{4Mr_e^2}\,.
\end{align}
Further notations used in literature are:
\begin{subequations}\label{106}
\begin{align}\label{106a}
    &\theta_{\alpha\alpha}=A+B+C&&\text{(trace of
the inertia moment tensor),}\\
&\mathcal{H}=\frac{C-\frac{1}{2}\,(A+B)}{C}&&\text{(dynamic
oblateness).}
\end{align}
\end{subequations}

\subsection{Rotation-symmetric specialization}

Rotation symmetry with respect to the $z$-axis means for the Newton
potential the condition
\begin{align}\label{107}
    \pderiv{\np}{\varphi}=0\,.
\end{align}
Hence for the Stokes coefficients results
\begin{align}\label{108}
    C_{nm}\,, \quad S_{nm}=0\qquad\text{for $n
= 2\,,\ 3\,, \dotsc$\;   ,}
\end{align}
i.e. from (\ref{91}) we find
\begin{align}\label{109}
    \np(r,\vartheta)=-\frac{\gn M}{r}
    \biggl[1-\sum\limits_{n=2}^{\infty}\biggl(\frac{r_0}{r}\biggr)^n
    J_{n}P_{n}(\cos\vartheta)\biggr]
\end{align}
($r_e\rightarrow r_0$ equatorial radius of the rotation-symmetric
body).

Passing over to the second order approximation, we get from
(\ref{109}) the expression
\begin{align}\label{110}
    \np(r,\vartheta)=-\frac{\gn M}{r}+\frac{\gn M
    r^2_0J_2}{4r^3}\,(1+3\cos (2\vartheta))\,.
\end{align}
With respect to the mass distribution the rotation symmetry means
the conditions
\begin{align}\label{111}
    \abc{a}m_{xx}=m_{yy}\,,\quad\abc{b}m_{xy}=m_{yz}=m_{zx}=0\,,\quad
    \text{i.e.}\quad\abc{c}m_{zz}=-2m_{xx}=-2m_{yy}\,.
\end{align}
Now we define the mass quadrupole moment, taking following sign
convention:
\begin{align}\label{112}
    m^{(Q)}=-m_{zz}\,.
\end{align}
Then, using spherical polar coordinates the formula (98) goes over
to
\begin{subequations}\label{113}
\begin{align}\label{113a}
   &
    \np(r,\vartheta)=-\frac{\gn M}{r}+\frac{\gn
    m^{(Q)}}{8r^3}\,\bigl(1+3\cos (2\vartheta)\bigr)\quad
    \text{with}\\ &
    r^2=R^2+z^2\,, \quad R=r\sin\vartheta\,, \quad
    z=r\cos\vartheta\,.
\end{align}
\end{subequations}
Here we would like to mention that the definition of some basic
notions in this field of research is rather confusing (choice of
different prefactors and signs), caused by both the two different
historical approaches: nuclear physics (see Landau/Lifschitz series
of textbooks) and geophysical and general-relativistic research (see
Misner, Thorne and Wheeler 1973).

In the rotation-symmetric case applied here the following formulas
hold:
\begin{align}\label{114}
    \abc{a}
    \theta_{11}=\theta_{22}=-\frac{1}{6}\,m^{(Q)}+\frac{2}{3}\,M_{\gamma\gamma}\,,
    \quad\abc{b}
    \theta_{33}=\frac{1}{3}\,m^{(Q)}+\frac{2}{3}\,M_{\gamma\gamma}\,,
    \quad\abc{c}
    \theta_{12}=\theta_{23}=\theta_{31}=0\,.
    \end{align}
We mention that in this context a basic role plays the polar moment
of inertia
\begin{align}\label{115}
    C=\theta_{33}=\int\limits_V\mu(\bar{\bm{r}})\nb{R}^2\dd
    \nb{V}\qquad
    (\nb{R}^2=\bar{x}^2+\bar{y}^2)\,,
\end{align}
Furthermore, we notice that in the notation (\ref{104}) the formulas
(\ref{114}a), (\ref{114}b), (\ref{105}) and (\ref{106}) read:
\begin{align}\label{116}
    \begin{aligned}
      &\abc{a}A=B=-\frac{1}{6}\,m^{(Q)}+\frac{2}{3}\,M_{\gamma\gamma}\,,
      &&\abc{b}C=\frac{1}{3}\,m^{(Q)}+\frac{2}{3}\,M_{\gamma\gamma}\,,
      &&\abc{c}J_{2}=\frac{C-A}{Mr_0^2}\,,\\
      &\abc{d}C_{22}=0\,,
      &&\abc{e}\theta_{\alpha\alpha}=2A+C\,,
      &&\abc{f}\mathcal{H}=\frac{C-A}{C}\,,
    \end{aligned}
\end{align}
whereas the comparison of (\ref{113}) with (\ref{110}) gives
\begin{align}\label{117}
    J_2=\frac{m^{(Q)}}{2Mr_0^2}=\frac{C-A}{Mr_0^2}\,.
\end{align}
Further results are:
\begin{align}\label{118}
    \abc{a}M_{\gamma\gamma}=A+\frac{C}{2}\,,\quad
    \abc{b}m^{(Q)}=2(C-A)\,.
\end{align}
Applying the above outcome to a rotation-symmetric ellipsoidal body
with the semi-axes $\hat{a}$, $\hat{c}$, ($\hat{a}>\hat{c}$) we
define the geometrical oblateness as usual:
\begin{align}\label{119}
    \Delta_{\phi}=\frac{\hat{a}-\hat{c}}{\hat{a}}
\end{align}
Let us now consider this body stationarily rotating about the
symmetry axis, where the interior centrifugal potential is given by
\begin{align}\label{120}
    \phi_{cf}(r,\vartheta)=-\frac{1}{2}\,\omega^2R^2=-\frac{1}{2}\,\omega^2r^2\sin^2\vartheta
    =
    \frac{1}{3}\,\omega^2r^2\bigl[1-P_2(\cos\vartheta)\bigr]
\end{align}
($\omega$ angular velocity, $P_2$ Legendre polynomial).

At the equator ($e$), where $r=\hat{a}$, $\vartheta=\dfrac{\pi}{2}$,
$P_2(\cos\dfrac{\pi}{2})=P_2(0)=-\dfrac{1}{2}$ is valid, results
\begin{align}\label{121}
    \phi_{cf|e}=-\frac{1}{2}\, \omega^2\hat{a}{}^2\,.
\end{align}
On the other hand the Newton potential (\ref{110}) reads on the
equator
\begin{align}\label{122}
    \phi_{\mathrm{N}|e}=-\frac{\gn M}{\hat{a}}-\frac{\gn
    M}{2\hat{a}}\,J_2\,.
\end{align}
Superposition of both the potentials gives the constant total
potential on the equator
\begin{align}\label{123}
    \phi_{tot|e}=\phi_{\mathrm{N}|e}+\phi_{cf|e}=-\frac{\gn
    M}{\hat{a}}\Bigl(1+\frac{1}{2}\,J_{2}+\frac{\omega^2\hat{a}{}^3}{2\gn
    M}\Bigr)=const\,,
\end{align}
since because of stability the total potential on the surface of the
ellipsoid has to be constant.

Analogous considerations with respect to the north pole ($p$), where
$r=\hat{c}$, $\vartheta=0$, $P_2(\cos 0)=P_2(1)=1$      holds, lead
to
\begin{align}\label{124}
    \phi_{tot|p}=-\frac{\gn
    M}{\hat{c}}\Bigr[1-\Bigr(\frac{\hat{a}}{\hat{c}}\Bigr)^2J_2\Bigr]\,.
\end{align}
Equating the results (\ref{123}) and (\ref{124}), we find
\begin{align}\label{125}
    J_2=\Bigl(\frac{\hat{c}}{\hat{a}}\Bigr)^2
    \frac{1-\dfrac{\hat{c}}{\hat{a}}-\dfrac{\omega^2\hat{c}{}^3}{2\gn
    M}}{1+\dfrac{1}{2}\Bigl(\dfrac{\hat{c}}{\hat{a}}\Bigr)^3}\,.
\end{align}
Using the approximation
\begin{align}\label{126}
    \abc{a}\Delta_{\phi}\ll 1\,,\quad
    \abc{b}\frac{\omega^2\hat{c}{}^3\Delta_{\phi}}{2\gn M}\ll 1\,,
\end{align}
we arrive at the interesting relationship (see Sharkow 1976)
\begin{align}\label{127}
    \Delta_{\phi}=\frac{3}{2}\,J_2+\frac{\omega^2\hat{a}{}^3}{2\gn
    M}
\end{align}
between the geometrical oblateness $\Delta_{\phi}$ and the dynamic
form factor $J_2$.

\section{Tidal and non-tidal influence on the rotation of the central body}
\subsection{Splitting of the angular velocity and angular acceleration of the rotating body}

The angular velocity of the central body $\omega_E$ may split into
the constant undisturbed angular velocity $\omega_{E|0}$, the tidal
angular velocity $\omega_{E|t}$ and the non-tidal angular velocity
$\omega_{E|nt}$:
\begin{align}\label{128}
    \omega_{E}=\omega_{E|0}+\omega_{E|t}+\omega_{E|nt}\,.
\end{align}
Temporal differentiation gives the angular acceleration in the form
\begin{align}\label{129}
   \dot{ \omega}_{E}=\dot{\omega}_{E|t}+\dot{\omega}_{E|nt}\,,
\end{align}
where $\dot{\omega}_{E|t}$ is the negative tidal acceleration (tidal
deceleration because of the braking effect of the oceanic waters)
and $\dot{\omega}_{E|nt}$ the non-tidal acceleration which usually
is identified with the rebound effect mentioned above (positive
sign). Our concept has to be enlarged, since according to our aim we
also have to grasp the cosmological influence. Therefore we separate
as follows:
\begin{align}\label{130}
    \dot{\omega}_{E|nt}=\dot{\omega}_{E|expl}+\dot{\omega}_{E|cos}
\end{align}
The first term $\dot{\omega}_{E|expl}$ refers to the explicit time
dependence of the polar moment of inertia of the central body,
including the above mentioned rebound time dependence, but admitting
further time-dependent effects. We call this amount ``explicit
temporal effect'' which according to this interpretation splits as
follows:
\begin{align}\label{131}
    \dot{\omega}_{E|expl}=\dot{\omega}_{E|rb}+\dot{\omega}_{E|further}
\end{align}
The second term $\dot{\omega}_{E|further}$ keeps open the
investigation of further effects.

\subsection{Treatment of the explicit temporal effect}

First we start our investigations by temporal differentiation of
(\ref{117}):
\begin{align}\label{132}
    \dot{J}_2=\frac{1}{Mr_0^2}\,\Bigl[\dot{C}-\dot{A}-\frac{2}{r_0}\,(C-A)\dot{r}_{0}\Bigr]\,.
\end{align}
As shown by previous research, in good approximation the earth
behaves like an incompressible body, being interpreted as a material
with constant trace of the inertia moment tensor:
$\Bar{\theta}=\theta_{\alpha\alpha}=const$. Application of this idea
to (\ref{116}e) leads to the equation
\begin{align}\label{133}
    \dot{A}=-\frac{1}{2}\,\dot{C}\,.
\end{align}
Inserting into (\ref{132}) gives
\begin{align}\label{134}
    \dot{J}_2=\frac{3}{2Mr^2_0}\,\Bigl[\dot{C}-2\Bigl(C-\frac{1}{2}\,\Bar{\theta}\Bigl)\frac{\dot{r}_0}{r_0}\Bigr]\,.
\end{align}
If the inequality
\begin{align}\label{135}
    2\Bigl(C-\frac{1}{3}\,\Bar{\theta}\Bigl)\frac{\dot{r}_0}{r_0}\ll
    \dot{C}
\end{align}
holds, then from (\ref{134}) follows
\begin{align}\label{136}
    \dot{J}_2=\frac{3\dot{C}}{2Mr^2_0}\,.
\end{align}
Eliminating $\dot{C}$ by means of the relation (\ref{25}a) and
giving the resulting relation the form
\begin{align}\label{137}
    \abc{a}\dot{C}=\ci{C}+C\varSigma\quad\text{with}\quad\abc{b}
    \ci{\theta}=\pderiv{C}{t}\,,
\end{align}
leads to
\begin{align}\label{138}
    \dot{J}_2=\frac{3}{2Mr_0^2}\,(\ci{C}+C\varSigma)\,.
\end{align}
Let us remember that the tidal braking effect of the earth is an
extreme long-term process (billions of years), whereas the
post-glacial rebound phenomenon is, cosmologically considered, a
short-term process (some ten thousands of years). Therefore one is
tempted to treat (under the relatively constant long-term tidal
background) the rotating earth as a rather isolated system with an
approximately conservative angular momentum (similar to the
well-known pirouette effect):
\begin{align}\label{139}
    \pderiv{}{t}\bigl(C\omega_E\bigr)=\ci{C}\omega_E+C\ci{\omega}_E=0\,,
\end{align}
i.e.
\begin{align}\label{140}
    \ci{C}=-\frac{C\ci{\omega}_E}{{\omega}_E}\,.
\end{align}
Inserting this expression into (\ref{138}) leads to the formula
\begin{align}\label{141}
    \dot{J}_2=-\frac{3C}{2Mr^2_0}\,\Bigl(\frac{\ci{\omega}_E}{\omega_E}-\varSigma\Bigr)\,.
\end{align}
If one neglects the cosmological influence ($\varSigma$) and reduces
the sight only on the rebound effect, i.e.
$\ci{\omega}_E=\dot{\omega}_{E|rb}$, then from (\ref{141}) follows
\begin{align}\label{142}
    \dot{J}_{2|rb}=-\frac{3C\dot{\omega}_{E|rb}}{2Mr^2_0\omega_{E}}\,.
\end{align}
This formula, on a different basis derived, was recently numerically
with success tested for the non-tidal rebound effect of the earth
(Wu, Schuh and Bibo 2003).

Apart from the rebound effect, among the further effects being under
consideration the hypothetical expansion of the earth will be
shortly treated now. In contrast to the rebound effect with its
temporally decreasing moment of inertia (increasing angular
velocity), obviously an expanding earth would exhibit an increasing
moment of inertia (decreasing angular velocity similar to the
braking behavior of the earth). For simplicity we consider a
rotating homogeneous sphere with the moment of inertia
\begin{align}\label{143}
    C=\frac{2}{5}\,Mr_0^2\,.
\end{align}
Considering the expansion process ($ep$), by explicit temporal
differentiation we receive
\begin{align}\label{144}
    \ci{C}_{ep}=\frac{2C}{r_0}\,\Bigl(\pderiv{r_0}{t}\Bigr)_{ep}>0\,.
\end{align}
Inserting into (\ref{138}) gives the relation
\begin{align}\label{145}
    \dot{J}_{2|ep}=\frac{3C}{Mr_0^3}\,\Bigl[\Bigl(\pderiv{r_0}{t}\Bigr)_{ep}+\frac{r_0}{2}\,\Sigma\Bigr]\,.
\end{align}
Some years ago we hypothetically developed an expansion theory of
celestial bodies on the basis of our PUFT (Schmutzer 2000, Schmutzer
2004) which led us to the following result for the radial expansion
velocity:
\begin{subequations}\label{146}
    \begin{align}\label{146a}
    &\Bigl(\pderiv{r_0}{t}\Bigr)_{ep}=4E_Sr_0\Bar{\sigma}^{4}\Sigma&&\text{with}\\
    &E_S=\frac{1}{4}\,\Sigma_S \Bigl(\frac{\gn
    M}{\Bar{\sigma}F_0}\Bigr)^2&&\text{(scalaric cosmic
expansion factor),}\\
&\Sigma_S=\frac{f_S\alpha_c}{3c_Q}&&\text{(scalaric material
factor),}
\end{align}
\end{subequations}
where following quantities are involved:
$F_0=\Bar{\sigma}\left|\bm{r}\times\dot{\bm{r}}\right|$ modified
areal velocity of the orbiting body (here moon), $\alpha_c$ cubic
thermal coefficient of dilation of the earth, $c_Q$ specific heat of
the earth, $f_S$ heat consumption factor of the earth. Let us
emphasize that the model for the earth (and therefore also the
values occurring) is very rough. Nevertheless we pointed as an
example to such a further possibility of  explicit temporal effects.

Inserting (\ref{146}a) into (\ref{145}) leads us to the final
formula for the temporal change of the dynamic form factor:
\begin{align}\label{147}
    \dot{J}_{2|ep}=\frac{12CE_S\Bar{\sigma}^{4}\Sigma}{Mr_0^2}\,\Bigl(
    1+\frac{1}{8E_S\Bar{\sigma}^4}\Bigr)>0\,.
\end{align}

\section{Specialization of the above developed theory to the case of
the Newtonian mechanics}\label{sec:9}

In order to simplify our above results based on PUFT, for usual
application on the basis of Newtonian physics we have to refrain
from the cosmological effects (acting on the 2-body system) by
setting
\begin{align}\label{148}
    \Sigma=0\,.
\end{align}
Then the store of the basic specialized formulas (getting from those
listed above) reads:
\begin{align}\label{149}
    \begin{aligned}
     & \abc{a}\ddot{\bm{r}}+\frac{\gn
      M}{r^3}\,\bm{r}+\bm{e}_R\,g_{rad}+\bm{e}_{\Phi}\,g_{az}=0\,,\quad
      \abc{b}\ddot{R}-R\Omega^2+\frac{\gn M}{R^2}+g_{rad}=0\,,\\
      &\abc{c}\dot{\Omega}+\frac{2\dot{R}\Omega}{R}+\frac{1}{R}\,g_{az}=0\,;
    \end{aligned}
\end{align}
\begin{subequations}\label{150}
    \begin{align}\label{150a}
    &g_{az}=\frac{\theta_E}{R
    m_{red}}\Bigl(\dot{\omega}_{E}+\frac{1}{\theta_{E}}\,\ci{\theta}_E\omega_E\Bigr)=-(R\dot{\Omega}
    +2\Omega\dot{R})\,,\\
    &g_{rad}=\Bigl(2+\frac{R\dot{\Omega}}{\Omega\dot{R}}\Bigr)R\Omega^2+\frac{1}{2m_{red}}\,\ci{\theta}_{E}
    \omega_{E}^2\,,\\
    &\dot{\omega}_E=-\frac{m_{red}R^2\Omega}{\theta_E}\,\Bigl(\frac{\dot{\Omega}}{\Omega}+2\frac{\dot{R}}{R}\,\Bigr)-
    \frac{1}{\theta_E}\,\ci{\theta}_E\omega_E=0\,,\\
    &\ddot{R}+R\Omega^2+\frac{\gn
    M}{R^2}+R^2\Omega\frac{\dot{\Omega}}{\dot{R}}+
    \frac{1}{2m_{red}\dot{R}}\,\ci{\theta}_{E}\omega_E^2=0\,;
\end{align}
\end{subequations}
\begin{align}\label{151}
    \abc{a}L_{tot}=L_{orb}+L_{rot}\,,\quad\abc{b}L_{orb}=m_{red}R^{2}\Omega\,,
    \quad\abc{c}L_{rot}=\theta_E\omega_E\,;
\end{align}
\begin{align}\label{152}
    \abc{a}\deriv{L_{tot}}{t}=0\,,\quad\abc{b}\deriv{L_{orb}}{t}=M_{az}=m_{red}R^{2}\Omega\Bigl(
    \frac{\dot{\Omega}}{\Omega}+2\frac{\dot{R}}{R}\Bigr)\,;
    \quad\abc{c}\deriv{L_{rot}}{t}=\theta_E\dot{\omega}_{E}\,;
\end{align}
\begin{align}\label{153}
    \abc{a}M_{az}=F_{az}R=-m_{red}Rg_{az}\,,\quad\abc{b}F_{az}=-m_{red}g_{az}\,;
\end{align}
\begin{align}\label{154}
    \begin{aligned}
      &\abc{a}E_{tot}=E_{orb|grav}+T_{rot}\,, &&\abc{b}
      E_{orb|grav}=T_{orb}+U_{grav}\,,&&\abc{c}T_{orb}=\frac{1}{2}\,m_{red}(\dot{R}^2+R^2\Omega^2)\,,\\
      &\abc{d}U_{grav}=-\frac{\gn M m_{red}}{R}\,,
      &&\abc{e}T_{rot}=\frac{1}{2}\,\theta_E\omega_E^2\,;
    \end{aligned}
\end{align}
\begin{align}\label{155}
    \begin{aligned}
     &
     \abc{a}\deriv{E_{rot}}{t}=\deriv{E_{orb|grav}}{t}+\deriv{T_{rot}}{t}\,,\\
     &
     \abc{b}\deriv{E_{orb|grav}}{t}=-\frac{1}{2}\,\ci{\theta}_E\omega_E^2\,,\quad\text{i.e.}\quad E_{orb|grav}
     =-m_{red}\Lambda+\frac{1}{2}\,\Hat{C}m_{red}\,,\\
     &\abc{c}
     \deriv{T_{rot}}{t}=\deriv{Q_{fric}}{t}+\frac{1}{2}\,\ci{\theta}_E\omega_E^2\,,\quad
     \abc{d}\deriv{Q_{fric}}{t}=\theta_E\omega_E\dot{\omega}_E<0\,,\quad
     \abc{e}\deriv{E_{tot}}{t}=\deriv{Q_{fric}}{t}<0\,;
    \end{aligned}
\end{align}
\begin{align}\label{156}
    \abc{a}E_{orb|grav}=-\frac{\gn M m_{red}}{2a}\,,
    \quad\abc{b}L_{tot}=J_{0}=m_{red}R^{2}\Omega+\theta_E\omega_E\,;
\end{align}
\begin{align}\label{157}
    \frac{1}{\Omega_{ell}}\,\deriv{\Omega_{ell}}{t}=-\frac{3}{2a}\,\deriv{a}{t}\,.
\end{align}

\section{Numerical evaluation of the theory}
\subsection{Main empirical values of physical and astrophysical quantities needed (adapted to the 2-body system)}
\begin{align}\label{158}
    &\gn=6.673\pot{-8}\kk{g^{-1}\,cm^3\,s^{-2}}&&
    \text{(Newtonian gravitational constant).}
\end{align}\pagebreak[3]
\textbf{\textit{Earth}} (for new values see Morrison and Stephenson
2001, Sabadini and Vermeersen 2004):
\begin{subequations}\label{159}
\begin{align}\label{159a}
&r_{E}=6.378\pot{8}\kk{cm} &&\text{(equatorial radius),}\\
&m_{E}=5.976\pot{27}\kk{g} &&\text{(mass),} \\
&\omega_E=7.292\pot{-5}\kk{s^{-1}}&&\text{(angular velocity),}\\
&\dot{\omega}_{E|t}=-6.15\pot{-22}\kk{s^{-2}}&&\text{(tidal angular
acceleration),}\\
&\dot{\omega}_{E|nt}=1.35\pot{-22}\kk{s^{-2}}&&\text{(non-tidal
angular acceleration),}\\
&\dot{\omega}_{E|emp}=\dot{\omega}_{E|t}+\dot{\omega}_{E|nt}=-4.8\pot{-22}\kk{s^{-2}}
&&\text{(empirical angular acceleration),}\\
&C_{E}=8.0394\pot{44}\kk{g\,cm^{2}}&&\text{(polar moment of
inertia),}\\
&A_E=8.0131\pot{44}\kk{g\,cm^2}&&\text{(equatorial moment of
inertia),}\\
&\ci{C}_E=-1.2675\pot{27}\kk{g\,cm^2\,s^{-1}}&&\text{(derivative of
the polar moment of  inertia),}\\
&J_{2}=1.0826\pot{-3}&&\text{(dynamic form factor),}\\
&\dot{J}_{2|rb}=-1.014\pot{-18}\kk{s^{-1}}&&\text{(derivative of the
dynamic form factor).}
\end{align}
\end{subequations}
\textit{\textbf{Moon}} (for new values of radial velocity and
orbital angular velocity see Anderson et al. 2002, Williams et al.
2003):
\begin{subequations}\label{160}
\begin{align}\label{160a}
    &m_{M}=7.348\pot{25}\kk{g}&&\text{(mass),}\\
    &v_{M|orb}=1.023\pot{5}\kk{cm\,s^{-1}}&&\text{(orbital
    velocity),}\\
    &v_{M|rad}=1.201\pot{-7}\kk{cm\,s^{-1}}&&\text{(radial
    velocity),}
    \\
    &\dot{\Omega}_{M}=-1.251\pot{-23}\kk{s^{-2}}&&\text{(orbital angular velocity).}
\end{align}
\end{subequations}
\textbf{\textit{Earth-moon system:}}
\begin{subequations}\label{161}
    \begin{align}\label{161a}
    &R=3.844\pot{10}\kk{cm}&&\text{(distance of the centers of
    mass),}\\
    &M=m_E+m_M=6.047\pot{27}\kk{g}&&\text{(total mass),}\\
    &m_{red}=\frac{m_E+m_M}{M}=7.259\pot{25}\kk{g}&&\text{(reduced
    mass).}
\end{align}
\end{subequations}

\subsection{Convenient parameters for fitting the theory to the empirical values}

For simplifying this numerical fitting procedure we introduce 4
adapting parameters $\alpha\,,\beta\,,\gamma\,,\delta$ being
understandable from the following equations:
\begin{align}\label{162}
    \begin{aligned}
      &\abc{a}\varSigma=\alpha\pot{-20}\kk{s^{-1}}\,,&&
      \abc{b}C_E=\beta\pot{44}\kk{g\,cm^2}\,,\\&\abc{c}\dot{\omega}_{E|emp}=-\gamma\pot{-22}\kk{s^{-2}}\,,
      &&\abc{d}\ci{C}_E=-\delta\pot{27}\kk{g\,cm^2\,s^{-1}}\,.
    \end{aligned}
\end{align}
Since here is not printing space enough for a detailed presentation
of the calculations, we only may report that the parameters
\begin{align}\label{163}
    \abc{a}\alpha=10.612\,,\quad\abc{b}\beta=8.0394\,,\quad\abc{c}\gamma=4.8\,,\quad\abc{d}
    \delta=1.2675
\end{align}
in context with the formulas (\ref{162}) fulfill the decisive
equations listed above. The last three parameters directly coincide
with the empirical results (\ref{159}g), (\ref{159}f), (\ref{159}i).

The first parameter follows from the relation (\ref{55}). Inserting
this value into (\ref{162}a) leads to the present numerical value of
the empirically important cosmological quantity ``logarithmic
 scalaric world function'' introduced above, which describes
the PUFT-caused cosmological influence on the 2-body system
considered:
\begin{align}\label{164}
    \varSigma=1.06\pot{-19}\kk{s^{-1}}=3.35\pot{-12}\kk{y^{-1}}\,.
\end{align}
This result is for our PUFT insofar of basic nature, since this
quantity determines the cosmological expansion influence on many
universal cosmogonical and astrophysical processes. Recently by
applying our theory to a closed homogeneous and isotropic cosmology
without the cosmological term (using the empirical WMAP-values of
2003) we arrived at the result
\begin{align}\label{165}
    \varSigma=5.2\pot{-19}\kk{s^{-1}}=1.64\pot{-11}\kk{y^{-1}}\,.
\end{align}
The new value (\ref{164}) is nearly one order of magnitude smaller
than the previous value (\ref{165}). This reduction improves the
situation with respect to the empirical results of the temporal
change of the empirical gravitational constant (Schmutzer 2004).

Concluding this section let us make two further annotations:

With respect to appropriate future accuracy measurements in this
field of research there is good hope for an improvement of the
measured values by some orders of magnitude in the next years
(Williams et al. 2003, Turyshev et al. 2003).

In section \ref{sec:9} we specialized our PUFT results to Newton
mechanics. If one repeats the calculations on this basis with
$\varSigma=0$, then one is confronted with some small discrepancies
which may be caused by some lack in the accuracy of measurements
(i.e. Newton mechanics may be correct), or by the mentioned
improvement of the accuracy these discrepancies are true (i.e. a new
theory as e.g. PUFT has to be tested).

\subsection{Report on further numerical results}

Various angular momenta:
\begin{align}
&\begin{aligned}
  &\abc{a}L_{orb}=2.857\pot{41}\kk{g\,cm^2\,s^{-1}}\,,
    &&\abc{b}L_{rot}=5.862\pot{40}\kk{g\,cm^2\,s^{-1}}\,,\\
    &\abc{c}L_{tot}=J_0=3.441\pot{41}\kk{g\,cm^2\,s^{-1}}\,;
\end{aligned}\label{167}
\intertext{Further:}
&\begin{aligned}
&\abc{a}g_{rad}=-2.609\pot{-1}\kk{cm\,s^{-2}}\,,&&\abc{b}g_{az}=-1.692\pot{-13}\kk{cm\,s^{-2}}\,;
\end{aligned}
\\
\label{168} &\begin{aligned}
&\abc{a}g_{rad|\mathrm{LTh}}=-1.565\pot{-14}\kk{cm\,s^{-2}}\,,&&\abc{b}g_{az|\mathrm{LTh}}=
1.838\pot{-26}\kk{cm\,s^{-2}}\,;
\end{aligned}\\[1ex]
\label{169} &\begin{aligned}
    &\abc{a}F_{az}=1.228\pot{13}\kk{g\,cm^2\,s^{-2}}\,,&&\abc{b}M_{az}=4.721\pot{23}\kk{g\,cm^2\,s^{-2}}\,;
    \end{aligned}
    \\ \label{170}&\begin{aligned}
    &\abc{a}T_{orb}=3.798\pot{35}\kk{g\,cm^2\,s^{-2}}\,,&&\abc{b}U_{grav}=-7.62\pot{35}\kk{g\,cm\,s^{-2}}\,,
    \\
    &\abc{c}E_{orb|grav}=-3.822\pot{35}\kk{g\,cm^2\,s^{-2}}\,;
    \end{aligned}\\[1ex]
    \label{171}&\begin{aligned}
&\abc{a}T_{rot}=2.137\pot{36}\kk{g\,cm^2\,s^{-2}}\,, &&\abc{b}
E_{tot}=1.755\pot{36}\kk{g\,cm^2\,s^{-2}}\,;\end{aligned}\\[1ex]
\label{172} &\begin{aligned}
  &\abc{a}\dot{E}_{orb|grav}=3.329\pot{18}\kk{g\,cm^2\,s^{-3}}\,,
&&\abc{b}
\dot{T}_{rot}=-3.128\pot{19}\kk{g\,cm^2\,s^{-3}}\,,\\
&\abc{c}\dot{Q}_{fric}=-2.814\pot{19}\kk{g\,cm^2\,s^{-3}}\,,&&
\abc{d}\dot{E}_{tot}=-2.795\pot{19}\kk{g\,cm^2\,s^{-3}}\approx
\dot{Q}_{fric}\,;
\end{aligned}\\[1ex]
\label{173}&\begin{aligned}
    \abc{a}\deriv{\Omega_{circle}}{R_{circle}}+\frac{3\Omega_{circle}}{2R_{circle}}\approx
    0\,,\quad\text{i.e.}\quad\abc{b}
    \deriv{\Omega_{circle}}{R_{circle}}+2\frac{\Omega_{circle}}{R_{circle}}\neq
    0\,.
\end{aligned}
\end{align}
With respect to both these equations we remember that the first one
coming from (\ref{79}) is well fulfilled (to be seen in connection
with the conservation of the total angular momentum), whereas the
second similar one resulting from (\ref{80}) is not fulfilled
because of the non-conservation of the orbital angular momentum of
the 2-body system (in contrast to a situation without external
influence).

I would like to thank very much for scientific discussions P.
Brosche, M. Schneider, H. Schuh, U. Walzer and Bu Win, and further
particularly A. Gorbatsiewich for scientific discussions and
technical help.

\section*{References}

\begin{Literatur}
  \item
  Anderson, J.G., Lau, E.L., Turyshev. S.G., Laing, Ph.A., Nieto, M.M.: 2002,
arXiv:gr-qc/0107022
    \item
Brosche, P. , Schuh, H.:  1998, in: Surveys in Geophysics 19,
(p.417-430)
    \item
L\"ammerzahl, C., Neugebauer, G.: 2001, in: Gyros, Clocks,
Interferometers: Testing Relativistic Gravity in Space (p.31-51),
Springer Verlag Berlin
    \item
Lambeck, K.: 1980, The Earth`s Variable Rotation: Geophysical Causes
and Consequences,    Cambridge University Press, Cambridge
    \item
Misner, Ch.W. ,Thorne, K.S. ,Wheeler, J.A.: 1973, Gravitation, W. H.
Freeman and Company, San Francisco
    \item
Morrison, L.V, Stephenson, F.R.: 2001, Journal of Geodynamics 32,
(p.247-265)
    \item
Sabadini, R. , Vermeersen, B.: 2004, Global Dynamics of the Earth,
Kluwer Academic Publishers, Dordrecht-Boston-London
    \item
Schmutzer, E.: 1989, Grundlagen der Theoretischen Physik, Deutscher
Verlag der Wissenschaften, Berlin; 2005, Wiley-VCH, Weinheim (3rd
edition)
    \item
Schmutzer, E.: 2000, Astron. Nachr. 321, 227; 2001, Astron. Nachr.
322, 207; 2002, arXiv:astro-ph/0201007
    \item
Schmutzer, E.: 2004, Projektive Einheitliche Feldtheorie mit
Anwendungen in Kosmologie                             und
Astrophysik, Verlag Harri Deutsch, Frankfurt/Main
    \item
Schneider, M.: 1999, Himmelsmechanik (vol. 4), Spektrum der
Wissenschaften/Akademischer Verlag Heidelberg-Berlin
    \item
Schneider, M. :2003, Zeitschrift f\"ur Geod\"asie, Geoinformation
und Landmanagement 128, (p.326-332)
    \item
Sharkow, W.N.: 1976, Der innere Aufbau von Erde, Mond und Planeten,
Teubner Verlagsgesellschaft Leipzig
    \item
Turyshev, S.G., Williams, J.G. , Nordtvedt, K. , Shao, M., Murphy,
Th.W: 2003, arXiv:gr-qc/0311039
    \item
Williams, J.G, Turyshev, S.G., Murphy, Th.W.:  2003,
arXiv:gr-qc/0311021
    \item
Wu, B. , Schuh, H. , Bibo, P. : 2003, Journal of Geodynamics 36, (
p.515-521)
\end{Literatur}

\end{document}